# Superradiant Emission of Ultra-Bright Photon Pairs in Doppler-Broadened Atomic Ensemble

Yoon-Seok Lee, Sang Min Lee, Heonoh Kim, and Han Seb Moon[*]

*Department of Physics, Pusan National University, Busan 609-735, Korea*

**With a recent rising interest of single photon superradiance due to its potential usefulness for efficient collection of single photon from an atomic ensemble, bright and narrow photon pair source is a key component in realization of quantum communication and quantum computer based on coherent interaction between light and atomic ensemble. We report the superradiant emission of ultra-bright photon pairs with a coincidence counting rate per input power of 64,600 cps/mW via spontaneous four-wave mixing in a thermal vapour cell. The photon-pair generation rate is enhanced by the contribution of two-photon coherence of almost atomic velocity groups in the Doppler-broadened ladder-type atomic system. The quadratic proportionality of the probability of detecting a heralded single photon as a function of the optical depth clarifies that the ultra-brightness results from the superradiance. In addition, a single photon superradiant beating at a high optical depth of the atomic ensemble is observed for the first time.**

With the pioneering Dicke's work [1], a coherent and collective radiation of spontaneous emission from an atomic ensemble has been an important role, particularly in the area of quantum communication and quantum computer which is based on the interaction between light and atomic ensemble [2-4], including the generation of a narrowband photon pair [5-9] and the buildup of efficient quantum memory [10, 11].

The essential framework of collective radiation is based on coherent superposition of a single excitation of each atoms in a spatially distributed atomic ensemble, which results in the many-body symmetrical entangled state, i.e., the so-called Dicke-like state, with the characteristic phase factor; $\exp(i\vec{k}\cdot\vec{r}_j)$, where $\vec{k}$ is a wave-vector of incident field and $\vec{r}_j$ is the position of each atom in the ensemble. Especially, in the multi-level atomic system, this collective spin state provides the phase-matching condition for the cascade spontaneous emission [12]. Thus, to date it has been actively exploited to efficiently generate narrowband photon pairs via a write-read process and spontaneous four-wave mixing (SFWM) in highly dense cold atomic gases [5-9]. Furthermore, the interest of a superradiant emission of single photon from the atomic ensemble has been rising due to its potential usefulness of the efficient collection of single photon. Recently, the experimental observation of a single-photon superradiance has been made via the write-read process in cold atoms [13] and H. H. Jen has presented a theoretical description of the superradiant emission of photon pairs via four-wave mixing in a ladder-type atomic ensemble, showing that the cooperative Lamb shift is observable and can be manipulated to be used for frequency qubits [14].

Despite successful implementation of narrowband paired photon source in cold atomic system, the generation rate is too low to be practically useful and the experimental complexity for obtaining ultracold temperatures still limits their applicability. Thus, a variety of sophisticated efforts have been made toward achieving high generation rates and replacing ultracold atoms with thermal vapour; room temperature quantum device [15, 16]. A Doppler-broadened atomic ensemble has an additional degree of freedom, i.e., velocity of each moving atom. Therefore, we can expand the above description of the collective radiation into a velocity space, considering the coherent superposition of the atomic excitation between different velocity groups in the ensemble with phase factor; $\exp(i\delta_j t)$, where $\delta_j$ is a detuning of incident field due to Doppler-frequency shift. This concept allows a range of interesting features to be considered, such as quantum memory based on atomic frequency comb technology [11] and superradiant Doppler beats [17]. Recently, constructive interference between different velocity classes of radiating atomic dipoles in a pulsed four-wave mixing process of a Doppler-broadened Rydberg atomic ensemble, which leads to signal revival, has been reported [18].

In this article, we report on the superradiant emission of highly bright photon pairs via SFWM in a ladder-type [87]Rb warm vapour. A coherent two-photon excitation of a wide range of velocity groups in the Doppler-broadened atomic ensemble prompts the collectively enhanced radiation of photon pairs into the phase-matched direction, leading to high generation rate. The probability of detecting a heralded single photon is proportional to the square of the optical depth (OD), which implies that the bright emission is due to the superradiance. In addition, superradiant beating [17] of a localized single photon in idler mode is observed, for the first time, to the best of our knowledge; this implies coherent superposition of two-photon amplitudes from different velocity classes.

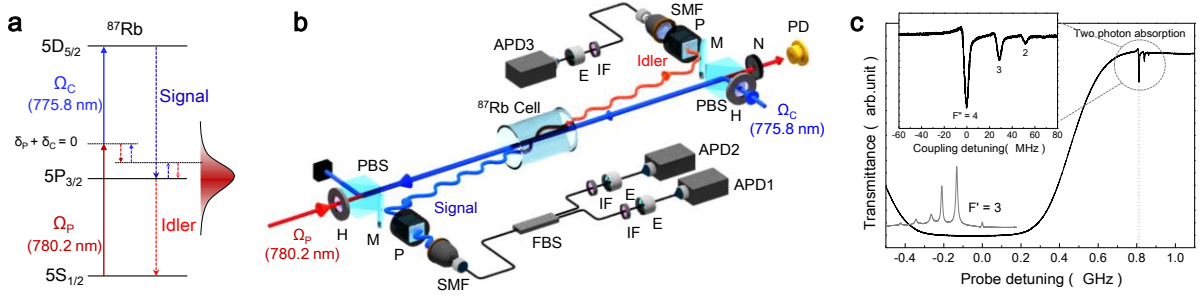

**Figure 1. Photon pair generation in a Doppler-broadened ladder-type atomic system. a**, Cascade emission of signal and idler photons via spontaneous four-wave mixing in Doppler-broadened three-level ladder-type atomic system interacting with pump and coupling fields, where $\Omega_{P,C}$ and $\delta_{P,C}$ are Rabi frequencies and detunings due to Doppler shift of pump and coupling fields, respectively. **b**, Schematic drawing of experimental set up. The counter-propagated pump ($\Omega_P$) and coupling ($\Omega_C$) fields make the coherent two-photon excitation in all around velocity groups, leading to photon pair generation in the phase-matched direction. The generated photon pairs are collected to two single mode fibers (SMF), respectively. P, polarizer; H, half-wave plate; PBS, polarizing beam splitter; PD, photodiode; N, neutral density filter; M, mirror; IF, interference filter; E, fused silica etalon filter; SMF, single mode fibre; FBS, fibre beam splitter; APD1-APD3, avalanche photodetectors. **c**, Pump-field transmittance spectra as a function of frequency detuning.

## Doppler-broadened ladder-type configuration

Figure 1**a** shows the configuration for the photon pair generation in the $5S_{1/2}$–$5P_{3/2}$–$5D_{5/2}$ transition of $^{87}$Rb of the Doppler-broadened atomic system. The pump ($\Omega_P$) and coupling ($\Omega_C$) fields are interacted with the $5S_{1/2}$–$5P_{3/2}$ and the $5P_{3/2}$–$5D_{5/2}$ transitions, respectively. $\delta_P$ and $\delta_C$ are the detuning frequencies from the resonances of the pump and coupling fields, respectively. The signal and idler photon pair can be generated in the $5S_{1/2}$–$5P_{3/2}$ and the $5P_{3/2}$–$5D_{5/2}$ transitions via SFWM, respectively. To make a large number of moving atoms coherently contribute to the generation of photon pairs in Doppler-broadened atomic system, strong two-photon coherence should be distributed throughout the almost atomic velocity groups, because two-photon coherence is an essential element for the coherent two-photon excitation, leading to the efficient generation of photon pairs via SFWM.

For SFWM process in a Doppler-broadened three-level ladder-type atomic system, two-photon amplitude is given by

$$\Psi_\upsilon(\tau) = A(\upsilon) e^{[-\Gamma_{31}/2 + i k_I \upsilon]\tau}, \qquad (1)$$

where $\Gamma_{31}$ is a decay rate of the $5S_{1/2}$–$5P_{3/2}$ transition, $k_I$ is wave-vector for an idler photon, $\upsilon$ is atom velocity, and $\tau$ is a detection time difference between signal and idler photons, according to first-order perturbation theory [19]. Here, $A(\upsilon)$ is a coefficient depending on the atom velocity, indicating the extent of two-photon coherence generated in each velocity group; the explicit formula is given in the **Supplementary Material** [20]. Especially in the $5S_{1/2}$–$5P_{3/2}$–$5D_{5/2}$ transition, $A(\upsilon)$ is not suppressed for the moving atoms but strongly exhibited because the almost velocity classes can satisfy the two-photon resonance condition under the counter-propagation configuration, as a result of the small difference in wavelength between the pump (780.2 nm) and coupling (775.8 nm) fields. Therefore, we can distribute the strong two-photon coherence over a wide range of velocity groups, which implies that effectively high OD was achieved.

The second-order correlation function for the paired photon generated from a Doppler-broadened atomic ensemble via SFWM is defined as

$$G^{(2)}_{SI}(\tau) = \left| \int \Psi_\upsilon(\tau) f(\upsilon) d\upsilon \right|^2, \qquad (2)$$

where $f(\upsilon)$ is a one-dimensional Maxwell-Boltzmann velocity distribution function. As the almost velocity groups in an ensemble can coherently contribute to photon-pair generation, $G^{(2)}_{SI}(\tau)$ is significantly enhanced via constructive interference between two-photon amplitudes from the different velocity groups. A more detailed theoretical description is presented in the **Supplementary Material** [20].

The experimental setup used in this study is shown in Fig. 1**b**. The pump ($\Omega_P$) and coupling ($\Omega_C$) laser fields counter-propagated through a 12.5 mm long vapour cell containing the $^{87}$Rb isotope and spatially overlapped completely with the same $1/e^2$ beam diameters of 1.2 mm. The pump ($\Omega_P$) and coupling ($\Omega_C$) field intensities were adjusted by using half-wave plate and polarized beam splitter and the temperature of a vapour cell was set to 52 ℃. The vapour cell was housed in three layers μ-metal chamber to prevent the decoherence, which would be induced by exposure to the external magnetic field. The signal and idler photons are generated via SFWM process in phase-matched direction and collected into two single mode fibres (SMF), respectively.

To verify that the two-photon coherence was strongly generated in the atomic ensemble, we monitored the transmittance spectra of $\Omega_P$, as shown in Fig. 1c. The narrow two-photon absorption is clearly apparent at a blue-detuned frequency far from the intermediate state, which means that two-photon coherence was strongly generated in the $5S_{1/2}$ (F = 2) – $5P_{3/2}$ (F' = 3) – $5D_{5/2}$ (F" = 4) transition of the atomic ensemble [21]. To maximize two-photon coherence and prevent the Rayleigh scattering; single absorbed and emitted fluorescence, the frequencies of both fields were stabilized at the 810 MHz blue-detuned two-photon absorption spectrum in the $5S_{1/2}$ (F = 2) - $5P_{3/2}$ (F' = 3) - $5D_{5/2}$ (F" = 4) transition out of the Doppler absorption profile. However, in the Doppler-broadened ladder-type configurations such as the $5S_{1/2}$ – $5P_{3/2}$ – $4D_{5/2}$ telecom-band transition and the $5S_{1/2}$ – $5P_{3/2}$ – $nD$ Rydberg transition of Rb atoms [15, 18], the wavelength difference between the pump and coupling fields are much larger than that of our configuration. Thus, in those ladder-type configurations, the narrow and strong two-photon absorption, such as Fig. 1c, cannot be observed, because of the two-photon Doppler shift due to the large wavelength difference. Therefore, the $5S_{1/2}$ (F = 2) – $5P_{3/2}$ (F' = 3) – $5D_{5/2}$ (F" = 4) transition is excellent configuration for the photon pair generation in a Doppler-broadened atomic system.

## Ultra-bright photon pair

We measured the signal and idler single counting rates ($N_S$ and $N_I$, respectively) and the net coincidence counting rate ($N_C$) as functions of the pump power for different coupling powers with a coincidence window of 4.1 ns, as shown in Fig. 2. Both single count results were corrected by considering the detector dead times of ~ 50 ns [22], whereas the $N_C$ were unchanged. All results were averaged over 10 measurements. At 48 mW coupling power, $N_C$ was 64,600 cps/mW, which is comparable to that of spontaneous parametric down-conversion (SPDC) in a $\chi^{(2)}$ nonlinear crystal with no other corrections such as filtering, coupling, and detection efficiency adjustments. To the best of our knowledge, this $N_C$ is the highest value for SFWM process in an atomic ensemble reported to date. At a low coupling power of 10 mW, where $N_C$ was 30,300 cps/mW out of the effect of dead time and saturation, both the single and coincidence counts exhibited good linearity as the pump power was increased to 1mW. However, a tendency towards saturation was observed, because of the limited atomic coherence and population of the ensemble under high power pumping, and multi-photon effect. From the experimental results, the number of generated photon pairs per input pump power $N_{pair}$ was estimated to be 8.98 ± 0.22 MHz/mW, from the equation $N_{pair} = (N_S \times N_I) / N_C$. This outstanding brightness is attributed to the collective enhancement of the emission rate due to the coherent contribution of the almost velocity classes in the Doppler-broaden atomic ensemble. The heralding efficiencies $\eta_{S,I} = N_C / N_{S,I}$ for both signal and idler photons were calculated as being 5.8 ± 0.1%.

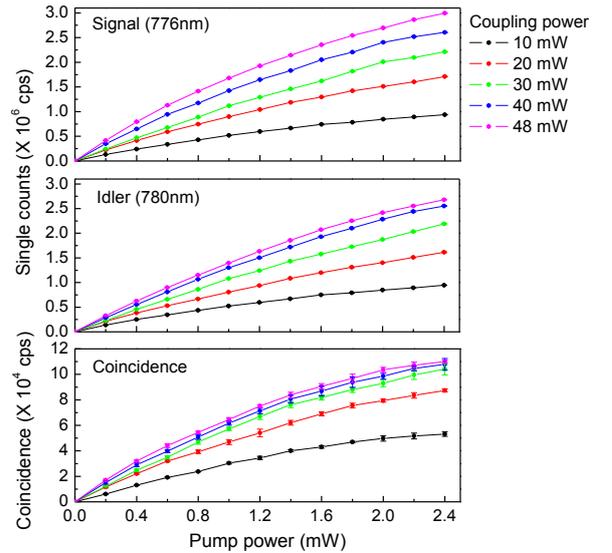

**Figure 2. Single and coincidence counting rate as functions of pump powers for the different coupling powers.** The single counting rates for signal and idler photons were corrected by considering the dead time of detectors. The coincidence counting rates were obtained by subtracting the accidental counting rates from the measured data with coincidence time window of 4.1 ns. All results were averaged over 10 measurements and the error bars are the standard deviation.

For a photon pair source reliant on a parametric process such as SPDC and SFWM, the strong temporal correlation between the signal and idler photons and the thermal nature of the individual photons can be observed via the Hanbury Brown-Twiss (HBT) experiment [23]. In our experiment, the temporal statistical properties of the photon pairs are presented in the **Supplementary Material** [20]. Under the condition of 1 mW pump and 10 mW coupling powers, the normalized cross-correlation function between the paired photons exhibited strong time correlation; $g^{(2)}_{SI}(0) = 84.7 \pm 0.01$, and the normalized auto-correlation for the individual signal and idler photons indicated the bunching properties of the thermal light; $g^{(2)}_{SS}(0) = 1.74 \pm 0.09$, and $g^{(2)}_{II}(0) = 1.74 \pm 0.06$. The correlation time of photon pairs was measured to $\tau_C = 1.87$ ns; the estimated bandwidth is approximately 540 MHz. The Cauchy-Schwarz inequality [24] for the second-order correlation function was strongly violated by a factor of 2370 ± 150 at $N_C$ = 30,300 cps/mW, which implies non-classical temporal correlation for paired photons. In addition, at 1 mW pump and 40 mW coupling powers, the normalized conditional auto-correlation function for the idler photons displays the same temporal shape as that of the cross-correlation, with a peak value of $g^{(2)}_{SII}(0) = 3.88 \pm 0.07$. From the measured temporal correlation functions, we found the standard second-order correlation for a heralded single photon in idler mode with clearly apparent anti-bunching, i.e., $g^{(2)}_{C}(0) = 0.138 \pm 0.003$ at $N_C$ = 64,600 cps/mW. By reducing the pump power (0.2 mW), we obtained $g^{(2)}_{C}(0) = 0.037 \pm 0.003$. We can thus confirm that the brightly emitted photon in idler mode was nearly a single photon with negligible multi-photon effect under the heralding detection of the signal photon.

### Superradiant emission

In this article, we experimentally demonstrated that our highly bright photon pairs are due to superradiant emission from a Doppler-broadened atomic ensemble. In particular, to confirm the superradient emission, we measured the quadratic increment of the correlated photon pairs to the number of atoms and the superradiant beats in a temporal histogram of signal and idler coincidence counts.

First, we note that the superradiance intensity is proportional to square of the particle number [1], whereas the intensity of the spontaneous radiation from individual atoms is linearly proportional to the number of atoms. Figure 3 shows $N_S$, $N_I$, and $N_C$ as functions of OD under the condition of 1 mW pump and 40 mW coupling power, where the OD is linearly proportional to the particle number [25]. We investigated the emission rates as functions of OD by adjusting the vapour cell temperature from 31 to 52 ℃. The OD values of the vapour cell at a given temperature were fitted to the measured linear absorption spectrum, using Beer's law; $OD \simeq \ln(I_t/I_0)$, where $I_t$ and $I_0$ are the transmitted and input intensity of pump beam, respectively [26].

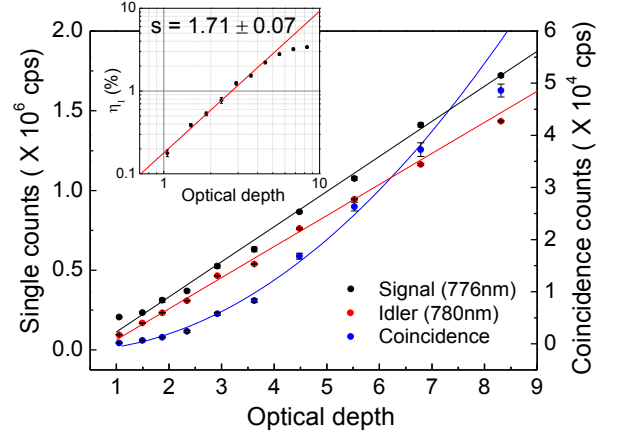

**Figure 3. Accelerated emission rate of photon pairs.** Single and coincidence counting rates for signal and idler photons are measured as functions of optical depth (OD). Black and red solid lines are linear fits and blue solid curve is a parabola fit. All fitting curves are based on the measurement until the OD is smaller than 7. Inset: Heralding efficiency as function of OD in log-log scale. The red line of inset graph is a linear fit with slope s = 1.71 ± 0.07.

The solid lines of Fig. 3 are linear (black and red lines) and quadratic (blue line) fitting curve to measured data up to OD < 7. The increment of $N_C$ is quadratic with increasing OD < 7, while that of individual single counting rates ($N_S$ and $N_I$) is linear. When the detection probability ($\eta_I$) of a heralded single photon in idler mode is described as the log-log graph for $\eta_I$ versus OD (Fig. 3, inset), we can see the accelerated emission of a heralded single photon, $\eta_I \propto OD^s$, with slope s = 1.71 ± 0.07. This observation is a representative result of the superradiance [1]. Note that the quadratic tendency is degraded at high OD > 7; this is primarily due to idler re-absorption in the atomic ensemble.

Second, the interesting oscillation appeared in the coincidence-event histogram within a time scale of several nanoseconds at high temperature, about 70 ~ 75 ℃, as shown in Fig. 4**a**. This observed oscillation is the beat between both two-photon amplitudes of the photon pairs generated from the separated two atomic velocity groups. The atomic ensemble is not only a photon emitter but also an effective photon absorber. After the idler photons are generated in the atomic ensemble, they can be significantly absorbed while passing through the highly dense atomic media, because the idler photons are resonant on the $5S_{1/2}$–$5P_{3/2}$ transition. As the absorption of idler photons becomes dominant at the resonance of the $5S_{1/2}$–$5P_{3/2}$ transition, the spectral feature of the idler photons coupled to the SMF can be considerably distorted from the Doppler-profile of a warm $^{87}$Rb vapour. Therefore at high OD, the remaining idler photons originate from the atomic velocity groups at far red- and blue-detuned wavelength, such as the insert spectrum of Fig. 4**b**. Further, as the two-photon amplitudes of different atomic velocity groups are coherently superposed, quantum interference occurs, leading to superradiant beating [17].

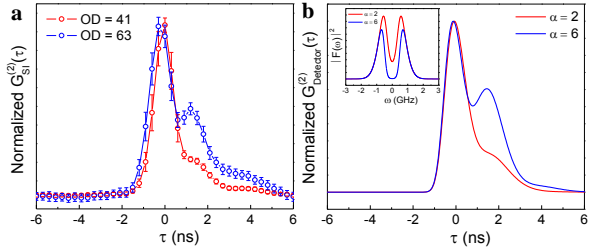

**Figure 4. Single photon superradiant beating. a**, Normalized temporal histogram of coincidence events at the OD = 41 and 63. **b**, Plots of normalized $G^{(2)}_{Detector}(\tau)$ considering corresponding filtering functions. Inset: Plots of $|F(\omega)|^2$ in the case of the α = 2 and 6.

To theoretically analyse the experimental results shown in Fig. 4**a**, we present a model that the observed oscillation results from the interference between wavelength- separated two-photon amplitudes due to the reabsorption filtering effect. Considering that the idler photons with frequencies within the Doppler bandwidth are filtered out after the generation process, the filtering function is constructed as [27]

$$F(\omega) = T(\omega)\exp[-\alpha A(\omega)], \quad (3)$$

where α is the absorption coefficient of $F(\omega)$, $T(\omega)$ are a Gaussian function applying to the etalon filter with 940 MHz bandwidth, and $A(\omega)$ is the Gaussian function of the Doppler-absorption profile with 540 MHz bandwidth. A plot of $|F(\omega)|^2$ is shown in the inset of Fig. 4**b**, for α = 2 and 6. Note that α is not the OD under these experimental conditions but, rather, a calculation parameter. As the filtering effect on the electric field operator in the time domain [28] can be expressed by

$$\hat{E}^{\dagger}_{Detector}(t) = \int dt' \hat{F}(t-t')\hat{E}^{\dagger}(t'), \quad (4)$$

where $\hat{E}^{\dagger}(t)$ is the electric field operator at time $t$ and $\tilde{F}(t)$ is the Fourier transform of $F(\omega)$, the final second-order correlation function on detector $G^{(2)}_{Detector}(\tau)$ can be obtained through simple convolution of $G^{(2)}_{SI}(\tau)$ and $|\tilde{F}(t)|^2$, such that

$$G^{(2)}_{Detector}(t) = \int d\tau |\tilde{F}(t-\tau)|^2 G^{(2)}_{SI}(\tau). \quad (5)$$

The calculation results are plotted in Fig. 4**b**, and exhibit good qualitative agreement with the experimental results. Therefore, the oscillation feature is due to interference between elements of different frequencies, i.e., the right- and left- hand sides of the strong absorption dip. Although this phenomenon has already been observed in the fluorescence from Doppler-broadened media, to the best of our knowledge, this is the first time superradiant beating for a single photon has been observed.

## Conclusions

We have demonstrated superradiant emission of ultra-bright photon pairs at wavelengths of 780.2 nm and 775.8 nm via spontaneous four-wave mixing from a Doppler-broadened atomic ensemble. The strong two-photon coherence was distributed throughout the almost velocity classes by using the counter-propagating geometry in the $5S_{1/2}$–$5P_{3/2}$–$5D_{5/2}$ transition of $^{87}$Rb and the collective enhancement of the radiation of photon pairs in the phase-matched direction from the Doppler-broadened atomic ensemble was exploited. Hence, highly bright photon-pair generation was obtained with a coincidence counting rate per input pump power of 64,600 cps/mW at a 48 mW coupling power and a relatively long coherence time of 1.87 ns. We revealed that the physical foundation for the outstanding performance of our system lies in the superradiant emission of photon pairs, and demonstrated the quadratic increment of the detection probability of heralded single photon as a function of the OD. The first observation of the single photon superradiant beating under high OD conditions indicates the coherent superposition and interference between two-photon amplitudes belonging to different velocity classes. This robust photon pair source has a high generation rate and relatively narrow bandwidth, which may be considered as a major resource for quantum optics. We believe that with the combination of fabrication technologies, such as microcell and hollow-core photonic crystal, our scheme can be one of the most applicable and scalable quantum devices towards realization of quantum communication, where quantum entanglement swapping between completely autonomous sources is essentially required.

## Methods

The generated photon pairs in phase-matched direction via SFWM from the $^{87}$Rb vapour cell were coupled to two SMFs, respectively. In order to isolate the signal and idler photons from the uncorrelated fluorescence, filtering boxes containing an interference filter with 3 nm the bandwidth and 95% transmittance and a solid fused-silica etalon filter (950 MHz full width at half maximum (FWHM) linewidth, 85% peak transmission) were employed. After passing through the filtering boxes, the photons were then coupled into multi-mode fibres. The estimated total efficiency for obtaining the spatially single mode and noise suppressed photon pairs was approximately 50%. The photons were detected by silicon avalanche photo-detectors (APD, PerkinElmer SPCM-AQRH-13HC; quantum efficiency ~40%, detection time jitter ~300 ps, dead time ~50 ns, measured dark count rates 100-300 Hz). To characterize the statistical properties of the photon pair source, measurement bench for the HBT experiment was set up. The apparatus included a fibre beam splitter and time-correlated single photon counting module (TCSPC, Picoharp 300) in start-stop mode with a 4 ps time resolution.

## Acknowledgements

This work was supported by the National Research Foundation of Korea (NRF) grant funded by the Korea government (MSIP) (No. 2015R1A2A1A05001819).